\documentclass[aip,reprint]{revtex4-1}

\usepackage{graphicx}
\usepackage{float}
\usepackage{tabularx}
\usepackage{adjustbox}
\usepackage{lipsum}
\usepackage{bbm}
\usepackage[dvipsnames]{xcolor,colortbl}
\definecolor{custom}{RGB}{0,144,226}
\usepackage{amsmath,mathrsfs,lineno,amssymb,amsbsy,setspace}

\draft 

\newcommand{\Revision}[1]{{\color{black} #1}} 

\begin{document}

\title{A blueprint for truncation resonance placement in elastic diatomic lattices with unit cell asymmetry} 



\author{Hasan B. Al Ba'ba'a}
\thanks{These authors contributed equally}
\altaffiliation{}
\affiliation{Department of Mechanical Engineering, Union College, Schenectady, NY 12308}
\affiliation{Department of Mechanical and Aerospace Engineering, University at Buffalo (SUNY), Buffalo, NY 14260 \looseness=-1}

\author{Hosam Yousef}
\thanks{These authors contributed equally}
\altaffiliation{}
\affiliation{Department of Mechanical and Aerospace Engineering, University at Buffalo (SUNY), Buffalo, NY 14260 \looseness=-1}

\author{Mostafa Nouh}
\email[Corresponding author: mnouh@buffalo.edu]{}
\affiliation{Department of Mechanical and Aerospace Engineering, University at Buffalo (SUNY), Buffalo, NY 14260 \looseness=-1}
\affiliation{Department of Civil, Structural and Environmental Engineering, University at Buffalo (SUNY), Buffalo, NY 14260 \looseness=-1}


\date{\today}

\begin{abstract}
\Revision{Elastic periodic lattices act as mechanical filters of incident vibrations. By and large, they forbid wave propagation within bandgaps and resonate outside them. However, they often encounter ``truncation resonances'' (TRs) inside bandgaps when certain conditions are met. In this study, we show that the extent of unit cell asymmetry, its mass and stiffness contrasts, and the boundary conditions all play a role in the TR location and wave profile. The work is experimentally supported via two examples that validate the methodology, and a set of design charts is provided as a blueprint for selective TR placement in diatomic lattices.}
\end{abstract}

\pacs{}

\maketitle 

\Revision{\textbf{Introduction.}} At their very core, elastic periodic lattices are effective mechanical filters of incident excitations. Lattices comprised of a sequence of two or more alternating layers exhibit Bragg bandgaps preventing vibrations within selected frequency ranges from being transmitted through the medium \cite{sigalas1992elastic,kushwaha1993acoustic}. These lattices represent a class of phononic crystals, owing to the similarity between the periodicity of lumped spring-mass elements and the ordered arrangement of atoms in a crystalline material \cite{Hussein2014,vasileiadis2021progress}. The majority of structural resonances associated with such lattices fall within their propagation frequencies or passbands. Yet, finite lattices are known to encounter natural frequencies that can avert these passbands and end up inside bandgaps, forcing a unique combination of high-amplitude deformations and an exponential spatial attenuation of the incident wave towards the bulk of the medium \cite{AlBabaa2016a}. Among other names, these natural frequencies have been referred to as in-gap or truncation resonances (TRs). The word ``truncation'' is motivated by the fact that these resonances do not appear in the dispersion diagram of an infinitely-repeating unit cell, but they result from the truncation of an infinite chain of unit cells to form a finite system with well-defined boundaries.

\vspace{0.1cm}

Localized modes associated with TRs provide a means to confine a large amount of energy to the edge of the structure. As a result, several applications have tapped into this phenomenon to achieve topological insulation \cite{susstrunk2015observation, Prodan2017DynamicalSystems, Rosa2019EdgeLattices}, flow control \cite{hussein2015flow, balasubramanian2022harnessing}, wave guidance \cite{lu2022defect, wu2009waveguiding}, defect characterization \cite{katch2023analysis}, and energy harvesting \cite{lv2019highly}. Despite this growing interest in TRs, studies of their existence criteria have been less prevalent. In 2011, Davis \textit{et al.}~examined anomalous TRs appearing in periodic flexural beams \cite {davis2011analysis}. Almost a decade later, Al Ba'ba'a \textit{et al.}~formalized the theory of TRs and established the role of unit cell symmetry in periodic rods \cite{al2023theory}. Park \textit{et al.}~extended this work to beams and provided supporting experiments \cite{park2024characteristics}. \Revision{Rosa \textit{et al.}~examined the topological origins of TRs in modulated structures via boundary phasons that shift the modulation profile while varying how the boundaries are truncated \cite{rosa2023material}.} Additional studies have shed light on different aspects of TRs, for example, those exhibited by locally resonant metamaterials \cite{sangiuliano2020influence}, or used to manipulate surface acoustic waves \cite{oudich2017phononic}. 

\Revision{Despite the presence of some analogies, elastic lattices of discrete atoms behave differently from continuous structures under similar conditions and therefore merit their own framework.} In 2017, Al Ba'ba'a \textit{et al.} derived the existence criteria of TRs in elastic lattices with two periodic atoms as well as uniform lattices with a periodic elastic foundation \cite{AlBabaa2017PoleDynamics}, and later generalized this analysis to polyatomic \cite{AlBabaa2019DispersionCrystals}. Bastawrous and Hussein extended these criteria to general boundary conditions as well as lattices with end masses \cite{bastawrous2022closed}. In this work, we present a blueprint for the design and selective placement of TRs in elastic periodic lattices. The paradigm presented here integrates all the pertinent factors that play a role in the onset of TRs, ranging from mass/stiffness contrasts and unit cell design to boundary effects, in order to delineate the theory of TRs in monatomic and diatomic lattices. Finally, we support our analytical framework with experimental measurements of the dynamic response of a finite diatomic spring-mass system, which validates the outlined theory.

\vspace{0.25cm}

\begin{figure*}[h!]
     \centering
\includegraphics[width=\linewidth]{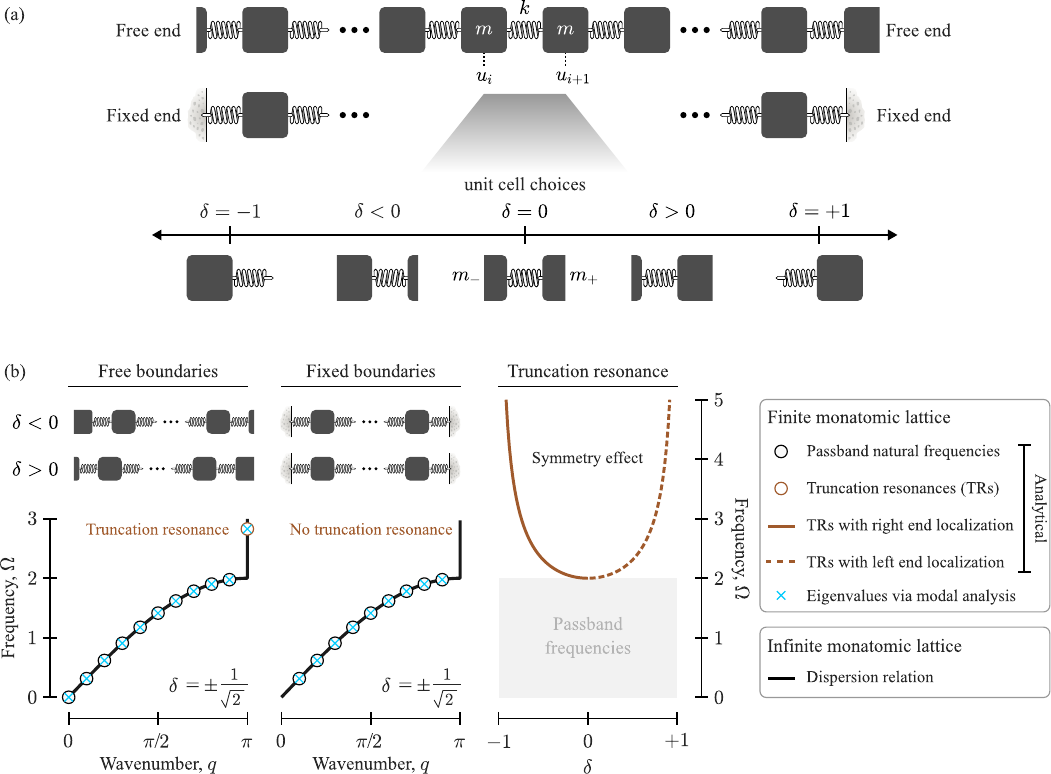}
     \begin{singlespace}
     \caption{\textbf{Monatomic lattice.} (a) Schematic diagram of a finite monatomic lattice comprised of a series of identical masses $m$ connected via identical springs $k$, subject to free or fixed boundary conditions.~Owing to its periodicity, the unit cell of the monatomic lattice can be defined in an infinite number of ways using the symmetry parameter $\delta$, where $-1 \leq \delta \leq 1$. While the unit-cell choice does not affect the dispersion relation, a finite lattice shall generally have different end masses whose values depend on the sign and value of $\delta$.~A positive $\delta$ will result in a heavier portion of the lumped mass ($> m/2$) on the right end of the lattice, and vice versa. $\delta = 0$ denotes a perfectly symmetric unit cell where $m_{+} = m_{-} = m/2$.~(b) Natural frequencies of a free-free (left) and fixed-fixed (middle) finite monatomic lattice of $n=10$ unit cells and $\delta = \pm 1/\sqrt{2}$, superimposed on the dispersion diagram of the infinite lattice.~TRs are marked with a circular marker of different color for distinction.~One TR appears in the free-free lattice at $\Omega = \sqrt{8}$ while none exist in the fixed-fixed case.~The rightmost plot tracks the location of the aforementioned TR as $\delta$ varies, showing perfect mirror symmetry about $\delta = 0$.~The line style depicts the lattice end associated with the TR's edge mode localization (\textit{solid}: right-end edge mode, $\delta <0$; \textit{dashed}: left-end edge mode, $\delta >0$).}
     \label{fig:Monatomiclattice}
     \end{singlespace}
\end{figure*}

\textbf{Monatomic lattice.} Consider a monatomic lattice represented by a chain of identical springs of stiffness $k$ and masses $m$ as shown in Fig.~\ref{fig:Monatomiclattice}(a), where $u_i$ denotes the displacement of the $i^{\text{th}}$ mass.~Owing to its periodicity, the building block of the infinite lattice (i.e., the unit cell) can be defined in an infinite number of ways. Using a symmetry parameter $\delta$, the lumped masses contained within a given unit cell can be parameterized as $m_\pm = \frac{m}{2}(1\pm\delta)$, where $m_\pm$ represents the masses connected to both ends of the spring $k$, as depicted in Fig.~\ref{fig:Monatomiclattice}(a). Consequently, the motion equations of an isolated unit cell are simplified to
\begin{subequations}
\begin{align}
    (1-\delta) \Ddot{u}_i + 2\omega_0^2(u_i - u_{i+1}) &= 2f_\text{L}/m \\
   (1+\delta) \Ddot{u}_{i+1} + 2\omega_0^2(u_{i+1} - u_{i}) &= 2f_\text{R}/m
\end{align}    
\end{subequations}
where $\Ddot{(~)}$ is an acceleration, $\omega_0 = \sqrt{k/m}$, and $f_{\text{L,R}}$ are the forces on both sides. Assuming a harmonic solution and casting these equations in a transfer matrix format~\cite{Hussein2014}, the state vectors at the $i^{\text{th}}$ and ${i+1}^{\text{th}}$ nodes of the lattice can be related via $\begin{Bmatrix} u_{i+1} & \epsilon_{i+1} \end{Bmatrix}^\text{T} = \mathbf{T} \begin{Bmatrix} u_{i} & \epsilon_{i} \end{Bmatrix}^\text{T}$, where $\epsilon_i = 2f_\text{L}/k$, $\epsilon_{i+1} = -2f_\text{R}/k$, and the transfer matrix $\mathbf{T}$ is given by
\begin{equation}
    \mathbf{T} = \begin{bmatrix}
    t_{11} & t_{12} \\
    t_{21} & t_{22}
    \end{bmatrix}=
    \begin{bmatrix}
    1-\frac{1}{2}(1-\delta)\Omega^2 & -\frac{1}{2} \\
    \frac{1}{2}\Omega^2(4-(1-\delta^2)\Omega^2) & 1-\frac{1}{2}(1+\delta)\Omega^2
    \label{eq:ML_TM}
    \end{bmatrix}
\end{equation}
where $\Omega = \omega/\omega_0$. The dispersion relation is defined as $\text{tr}(\mathbf{T}) = 2 \cos(q)$, which gives $\Omega = \sqrt{2(1-\cos(q))}$ with $q$ being a nondimensional wavenumber.~As expected, the dispersion relation is indifferent to the unit cell configuration, as evident by its independence of $\delta$~\cite{al2023theory}.

Once an infinite lattice is truncated to a finite size, $\delta$ plays an instrumental role in its behavior. The state vectors at the two lattice ends can be related via $\mathbf{T}^n$, where $n$ is the number of cells in the finite lattice.~An expression for $\mathbf{T}^n$ can be found as follows \cite{lin1969dynamics}
\begin{equation}
    \mathbf{T}^n = \frac{\cos(nq)}{2\cos(q)} \left(\mathbf{T} + \mathbf{T}^{-1} \right) + \frac{\sin(nq)}{2\sin(q)} \left(\mathbf{T} - \mathbf{T}^{-1} \right)
    \label{eq:Tn_1}
\end{equation}
in which the four matrix elements of $\mathbf{T}^n$ henceforth identified as $t_{11n}$ (top left), $t_{12n}$ (top right), $t_{21n}$ (bottom left), and $t_{22n}$ (bottom right) are given by
\begin{subequations}
\begin{align}
     \label{eq:t_11n} t_{11n} &= \cos(nq) + \frac{\sin(nq)}{2\sin(q)} \delta \Omega^2\\
     \label{eq:t_12n} t_{12n} &= -\frac{\sin(nq)}{2\sin(q)}\\
    \label{eq:t_21n} t_{21n} &= \Omega^2 \left(4-(1-\delta^2)\Omega^2 \right) \frac{\sin(nq)}{2\sin(q)}\\ 
     \label{eq:t_22n} t_{22n} &= \cos(nq) - \frac{\sin(nq)}{2\sin(q)} \delta \Omega^2
\end{align}
\label{eq:T_n_ML}
\end{subequations}

\Revision{In the absence of forces on both sides, the transfer matrix equation reduces to $t_{21n}=0$, which represents the characteristic equation of a free-free lattice} and results in three sets of natural frequencies: (i) $\Omega = 0$ (rigid body mode), (ii) $\Omega = 2 (1-\delta^2)^{-\frac{1}{2}}$, and (iii) a group of $n-1$ frequencies which can be computed by setting $ q = r \pi/n$ in the lattice's dispersion relation, with $r = 1,2, \dots, n-1$. Groups (i) and (iii) represent $n$ natural frequencies that fall within the propagation zone of the monatomic lattice and satisfy the dispersion relation. In other words, each of these frequencies corresponds to a mode shape generating an $[\Omega,q]$ point that lies on the lattice's dispersion curve, as shown in Fig.~\ref{fig:Monatomiclattice}(b). The natural frequency given by (ii), however, is larger than the cutoff frequency of the lattice at $\Omega = 2$ for any non-zero values of $\delta \in (-1,1)$.~It coincides with the cutoff frequency at $\delta = 0$ and rapidly approaches infinity as $|\delta|\rightarrow1$. This natural frequency indicates a unique resonance that lies outside the lattice's passband, and is therefore a TR. The leftmost panel of Fig.~\ref{fig:Monatomiclattice}(b) shows the dispersion diagram of a monatomic lattice with a unit cell configuration corresponding to $\delta = \pm 1/\sqrt{2}$. It consists of a single dispersive branch that ends at the edge of the irreducible Brillouin zone ($q = \pi$) and is followed by an unbounded stop band starting at the cutoff frequency ($\Omega = 2$). Shown in the same figure are the $n+1 = 11$ natural frequencies of a finite free-free monatomic lattice constructed from $n=10$ unit cells of the same configuration. As can be seen, a TR at $\Omega = \sqrt{8}$, as obtained from group (ii), can be spotted in the lattice's stop band. All natural frequencies are further confirmed via conventional modal analysis of the lattice's full mass and stiffness matrices. The numerically-obtained eleven eigenfrequencies (cross markers) perfectly match the analytical natural frequencies (circular markers) obtained from all three groups. Finally, it is noted that the mode shape associated with the TR shows a vibration localization at the edge with the `lighter' mass. As a result, a negative $\delta$ results in an energy localization at the right end of the monatomic lattice, while a positive $\delta$ induces localization at its left end, as captured in the rightmost plot of Fig.~\ref{fig:Monatomiclattice}(b).

The natural frequencies of a fixed-fixed finite lattice are effectively a subset of the free-free resonances, and are therefore fairly straightforward.~They are obtained by setting $t_{12n} = 0$, which gives a set of $n-1$ frequencies that are identical to those found in group (iii) of the free-free system. Notably, the fixed-fixed lattice has two natural frequencies less than its free-free counterpart due to the zero-displacement constraint on the masses at both ends, effectively eliminating two degrees of freedom from the dynamical model as depicted in Fig.~\ref{fig:Monatomiclattice}(a). One of the two eliminated natural frequencies is the rigid body mode at $\Omega=0$ which no longer exists once the lattice is anchored to fixed walls at both ends.~The other one is the TR which also no longer exists since a finite monatomic lattice is guaranteed to be mirror-symmetric about its center once both boundaries are fixed, yielding a response that is independent of $\delta$. The middle panel of Fig.~\ref{fig:Monatomiclattice}(b) shows the dispersion diagram and natural frequencies obtained for the same parameters used in the free-free case, namely $\delta = \pm 1/\sqrt{2}$ and $n=10$. The results confirm the lack of a TR in the fixed-fixed lattice. Since the dispersion curve is obtained from the unit cell analysis, it is independent of the finite lattice's boundary conditions as evidenced by the identical dispersion diagram obtained for both systems. 

\Revision{The two remaining scenarios, i.e., free-fixed and fixed-free lattices, are indicative of asymmetric boundary conditions and their characteristic equations are given by $t_{11n} = 0$ and $t_{22n} = 0$, respectively.~The fact that $t_{11n}(\delta) = t_{22n}(-\delta)$ shows that one of these cases can be obtained from the other by inverting the sign of the symmetry parameter, making it sufficient to study either one of them.} Let's start by examining a couple of special cases. If $\delta=0$ (perfectly symmetric unit cell), both characteristic equations become identical and reduce to $\cos(nq) = 0$, resulting in $n$ passband natural frequencies satisfying $q = \frac{2r-1}{2n}\pi$, with $r = 1,2, \dots, n$. If $\delta = \pm 1$, the free-end mass of the lattice is either $0$ or $m$, and the characteristic equations reduce to $\cos((2n\mp1)q/2) = 0$, again revealing the absence of TRs. However, the passband natural frequencies in this case satisfy $ q = \frac{2r-1}{2(n\mp1)}\pi$, with $r = 1,2,\dots,\varepsilon$, where $\varepsilon=n$ or $n-1$ depending on whether the free-end mass exists or not, respectively. For a free-fixed or fixed-free lattice corresponding to any other value of $\delta$ that is not equal to $0$ or $\pm 1$, a TR shall appear only if the free end contains the `lighter' mass, or in other words, a mass less than $m/2$. As such, a free-fixed lattice will only have a TR if $\delta>0$, while a fixed-free one will only have a TR if $\delta < 0$.

\vspace{0.3cm}

\begin{figure*}[h!]
     \centering
\includegraphics[width=\linewidth]{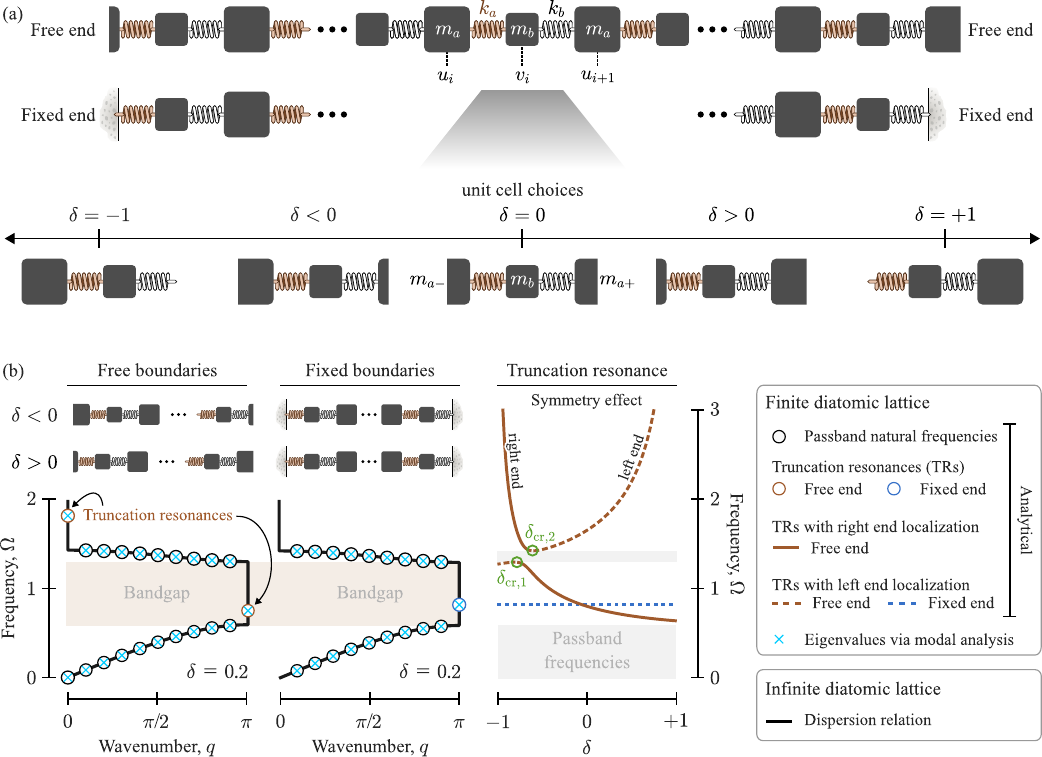}
    \begin{singlespace}
     \caption{\textbf{Diatomic lattice.} (a) Schematic diagram of a finite diatomic lattice comprised of a series of alternating masses $m_{a,b}$ connected via alternating springs $k_{a,b}$, subject to free or fixed boundary conditions.~Owing to its periodicity, the unit cell of the diatomic lattice can be defined in an infinite number of ways using the symmetry parameter $\delta$, where $-1 \leq \delta \leq 1$, which dictates the values of $m_{a+}$ and $m_{a-}$ (together forming the lumped mass $m_a$) that sit on either side of the intermediate mass $m_b$.~While the unit-cell choice does not affect the dispersion relation, a finite lattice shall generally have different end masses whose values depend on the sign and value of $\delta$.~A positive $\delta$ will result in a heavier portion of the lumped mass ($> m_a/2$) on the right end of the lattice, and vice versa. A zero value of $\delta$ denotes a perfectly symmetric unit cell where $m_{a+} = m_{a-} = m_a/2$. (b) Natural frequencies of free-free (left) and fixed-fixed (middle) finite diatomic lattices of $n=10$ unit cells, $\delta = 0.2$, $\varrho=1/3$, and $\vartheta = -3/5$, superimposed on the dispersion diagram of the infinite lattice. TRs are marked with a circular marker of different color for distinction. Two TRs appear in the free-free lattice at $\Omega = 0.74, 1.80$ while one exists in the fixed-fixed case at $\Omega = 0.82$. The rightmost plot tracks the location of the aforementioned TRs as $\delta$ varies. It shows that the frequency of the TR associated with a fixed boundary remains unchanged while that associated with a free boundary varies depending on the value of $\delta$. \Revision{The plot also identifies two critical values ($\delta_{\text{cr},1}=-0.81$ and $\delta_{\text{cr},2}=-0.60$) at which the TRs touch the lattice's passband and the edge mode shifts from one end to the other.}}
     \label{fig:DiatomicLattice}
     \end{singlespace}
\end{figure*}

\textbf{Diatomic lattice.} Consider a diatomic lattice where the spring-mass systems $[m_a, k_a]$ and $[m_b,k_b]$ alternate periodically, with $u_i$ and $v_i$ denoting the displacements of the $i^\text{th}$ $m_a$ and $m_b$ masses, respectively. Similar to the monatomic lattice, the unit cell configuration can be chosen based on $\delta$, as shown in Fig.~\ref{fig:DiatomicLattice}(a).~For any unit cell choice, the motion equations of an individual diatomic unit cell can be cast as a matrix differential equation: $ \mathbf{M}\Ddot{\mathbf{z}} + \mathbf{K}{\mathbf{z}} = \mathbf{f}$, where $\mathbf{z}^\text{T} = \begin{Bmatrix} {v}_i& {u}_i & {u}_{i+1} \end{Bmatrix}$, $\mathbf{f}^\text{T} = \begin{Bmatrix} 0& {f}_\text{L} & {f}_\text{R} \end{Bmatrix}$, $\mathbf{M}$ is a diagonal matrix of $\begin{Bmatrix} m_b & m_{a_-} & m_{a_+} \end{Bmatrix}$ where $m_{a_\pm} = m_a(1\pm\delta)/2$, and $\mathbf{K}$ is given by
\begin{equation}
\mathbf{K}=
    \begin{bmatrix}
    k_a+k_b &-k_a &-k_b \\
    -k_a & k_a & 0\\
    -k_b & 0 & k_b \\
    \end{bmatrix}
\end{equation}
To streamline the analysis, a normalizing frequency $\omega_0$, stiffness contrast $\vartheta$, mass contrast $\varrho$, and a $\gamma$ parameter that combines both contrasts are introduced as follows, respectively: 
\begin{subequations}
\begin{align}
        \omega_0 &= \sqrt{\frac{(k_a + k_b)(m_a + m_b)}{2 m_a m_b}} \label{eq:w0}\\
    \vartheta &= \frac{k_b - k_a}{k_b + k_a} \label{eq:vartheta}\\
    \varrho &= \frac{m_b - m_a}{m_b + m_a} \label{eq:varrho}\\
    \gamma &= (1-\vartheta^2)(1-\varrho^2)
\end{align}
\end{subequations}
Both $\vartheta$ and $\varrho$ can assume values between $-1$ and $1$, resulting in $\gamma \in [0,1]$.~Assuming a harmonic solution and dissolving the internal degree of freedom $v_i$ via Guyan reduction, a transfer matrix equation of the form $\begin{Bmatrix} u_{i+1} & \epsilon_{i+1} \end{Bmatrix}^\text{T} = \mathbf{T} \begin{Bmatrix} u_{i} & \epsilon_{i} \end{Bmatrix}^\text{T}$ is derived, where $\epsilon_i = f_\text{L}/(\omega_0^2 m_a)$ and $\epsilon_{i+1} = -f_\text{R}/(\omega_0^2 m_a)$, with the four components of $\mathbf{T}$ being
\begin{widetext}
\begin{subequations}
\begin{align}
    t_{11} &= 1-\frac{2}{\gamma}\Big[2-\delta(1-\varrho)-\vartheta(1+\varrho)\Big]\Omega^2 + \frac{2}{\gamma}(1-\delta)\Omega^4 \\
    \label{eq:t12_DL} t_{12} &= \frac{4}{\gamma}\Big[\Omega^2-(1-\varrho)\Big] \\
    \label{eq:t21_DL} t_{21} &= \frac{1}{\gamma} \Omega^2 \Big[2(1-\vartheta^2)(1+\varrho)-\big(4-(1-\varrho)(1+\delta^2) 
    -2\delta\vartheta(1+\varrho)\big)\Omega^2+(1-\delta^2)\Omega^4 \Big] \\
    t_{22} &= 1-\frac{2}{\gamma}\Big[2+\delta(1-\varrho)+\vartheta(1+\varrho)\Big]\Omega^2 + \frac{2}{\gamma}(1+\delta)\Omega^4
\end{align}
\label{eq:TMM_comp_DL}
\end{subequations}
\end{widetext}

Analogous to the analysis of the monatomic lattice, a dispersion relation for the diatomic lattice can be obtained, which has two branches described by $\Omega = \sqrt{1 \pm \sqrt{\smash[b]{1-\gamma \sin^2(\frac{q}{2})}}}$.~A Bragg bandgap opens up in between the two branches, the upper and lower limits of which ($\Omega_u$ and $\Omega_l$) correspond to the solutions of the branch equations at $q=\pi$, resulting in a bandgap width of $\Delta \Omega= \sqrt{2-2\sqrt{\gamma}}$. As such, smaller values of $\gamma$ result in larger bandgap widths and the bandgap remains open as long as $\gamma \neq 1$.~Finally, we note that the cutoff frequency of the diatomic lattice, given by the non-zero solution of the dispersion relation at $q = 0$, remains constant at $\Omega = \sqrt{2}$ regardless of the magnitude of $\gamma$. 

We focus the remainder of this discussion on the TRs of the finite diatomic lattice under different boundary conditions, with the understanding that passband natural frequencies can always be obtained from $\mathbf{T}^n$, following the same procedure outlined for the monatomic lattice. As previously established for solid continua, the TRs of a bi-layered phononic crystal with symmetric boundary conditions (e.g., free-free or fixed-fixed) coincide with the natural frequencies of the individual unit cell itself when subject to the same boundary conditions \cite{al2023theory}. As such, the TRs of the free-free diatomic lattice can be obtained from the roots of Eq.~(\ref{eq:t21_DL}), excluding $\Omega =0$, resulting in a biquadratic equation, from which the following observations can be made:
\begin{enumerate}
\itemsep0em 
    \item The equation has two positive roots culminating in two TRs: A lower frequency one inside the Bragg bandgap and a higher frequency one above the lattice's cutoff frequency (the latter lies exactly at the cutoff frequency if $\delta = \vartheta$).
    
    \item The equation becomes quadratic when $|\delta| = 1$, since the number of degrees of freedom of the finite lattice reduces by one. The higher frequency TR ceases to exist (mathematically, it moves to infinity similar to the monatomic lattice case). The lower frequency TR remains inside the bandgap unless $\varrho = 0$ (zero mass contrast) at which case it moves to one of the two bandgap limits, rendering the lattice free of any bandgap resonances.
    
    \item A perfectly symmetric unit cell ($\delta = 0$) combined with zero stiffness contrast ($\vartheta = 0$) moves the lower frequency TR to the bandgap limit located at $\Omega = \sqrt{1+\varrho}$, which could be either the lower or upper limit depending on the sign of $\varrho$.~Recall that $\delta = \vartheta$ forces the higher TR to coincide with the lattice's cutoff frequency.
\end{enumerate}

The leftmost panel of Fig.~\ref{fig:DiatomicLattice}(b) shows the dispersion diagram of a diatomic lattice corresponding to $\delta = 0.2$, and mass and stiffness contrasts of $\varrho = 1/3$ and $\vartheta = -3/5$, respectively. The lattice has two wave modes described by an acoustic (lower) branch and an optical (upper) branch which sit on both ends of the Bragg bandgap.~Shown in the same figure are the $2n+1 = 21$ natural frequencies of a finite free-free diatomic lattice constructed from $n=10$ unit cells of the same configuration, confirming the two TRs stated in observation \#1, one falling inside the bandgap at $\Omega = 0.74$ and the other taking place above the optical branch at $\Omega = 1.80$.

The fixed-fixed diatomic lattice has a single TR at $\Omega = \sqrt{1-\varrho}$, representing the positive root of Eq.~(\ref{eq:t12_DL}), and shows, once again, independence of the symmetry parameter $\delta$. The edge mode profile accompanying this TR exhibits localized displacement at the stiffer end of the lattice.~In other words, the localization takes place at the right end if $\vartheta >0$ and vice versa. Interestingly, in the case of $\varrho = 0$, the TR occurs at a constant frequency of $\Omega = 1$ regardless of the stiffness contrast $\vartheta$, a feature which has been exploited in diatomic lattices with topological protection \cite{chen2018topological}. The middle panel of Fig.~\ref{fig:DiatomicLattice}(b) displays the behavior of the fixed-fixed diatomic lattice and confirms the presence of a sole TR inside the bandgap at $\Omega = 0.82$ for the same values of $\varrho$ and $\vartheta$ used in the free-free example. Finally, given the dependence of both TRs of the free-free diatomic lattice on $\delta$, it is important to highlight this effect of unit cell symmetry on the edge mode profile associated with each of these two resonances, which is captured by the rightmost plot of Fig.~\ref{fig:DiatomicLattice}(b). \Revision{As can be seen, there exist two critical values $\delta_{\text{cr},1}$ and $\delta_{\text{cr},2}$ at which the TRs touch the lattice's passband, which can be obtained by substituting the $\Omega$ values corresponding to the upper bandgap limit and the cutoff frequency in $t_{21}=0$. Following that point, the end of the lattice which exhibits localized displacement shifts from one side to the other. The same plot shows a flat line tracking the location of the fixed-fixed TR with a varying $\delta$, confirming the lack of a symmetry effect on that particular truncation resonance.}

\begin{figure*}[ht]
     \centering
\includegraphics{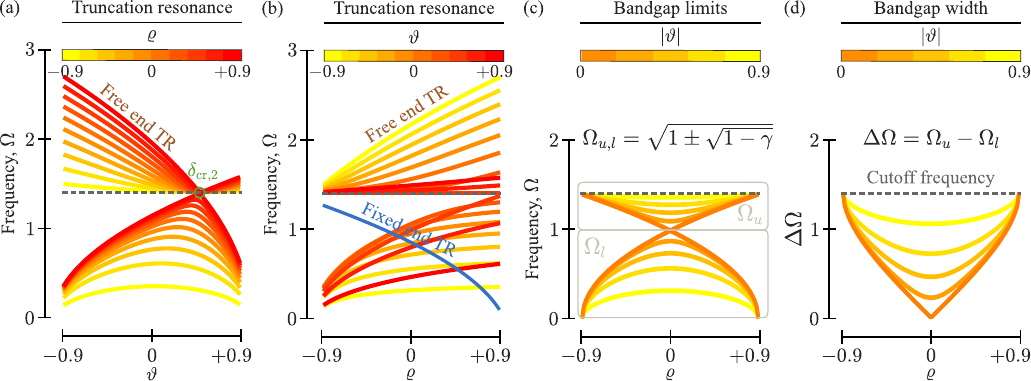}
    \begin{singlespace}
     \caption{\Revision{\textbf{TR design blueprint.} (a, b) Dual effect of $\varrho$ and $\vartheta$ on TR location in a diatomic lattice with a symmetry parameter of $\delta=0.5$. (c, d) Dual effect of $\varrho$ and $\vartheta$ on the upper and lower bandgap limits, $\Omega_u$ and $\Omega_l$, and the bandgap width $\Delta \Omega$. A bandgap closes at $\varrho = \vartheta = 0$ resulting in $\Delta \Omega=0$, indicative of a monatomic lattice.}}
     \label{fig:Parameters}
    \end{singlespace}
\end{figure*}

\Revision{Figure 3 summarizes the effect of the mass and stiffness contrasts on the truncation resonance, the bandgap limits, and the bandgap width, and provides a culminating blueprint for TR placement in diatomic lattices. A symmetry parameter of $\delta=0.5$ is chosen. The figure shows that a TR associated with a fixed boundary depends solely on $\varrho$, as inferred from Eq.~(\ref{eq:t12_DL}), and must fall below the lattice's cutoff frequency, whereas a TR associated with a free boundary is influenced by both $\varrho$ and $\vartheta$, and can either fall inside the Bragg bandgap or in the stop band above the cutoff frequency. The latter is shown in Fig.~\ref{fig:Parameters}(a) to coincide with the cutoff frequency when $\vartheta = \delta$, as stated in the first bullet point earlier, thus defining the $\delta_{\text{cr},2}$ point. Finally, Figs.~\ref{fig:Parameters}(c) and (d) highlight the effect of both $\varrho$ and $\vartheta$ on the Bragg bandgap limits and width, and show that the gap closes in the absence of both contrasts (i.e., when $\varrho = \vartheta = 0$), resulting in a monatomic lattice. The design charts in Figs.~\ref{fig:DiatomicLattice}(b) and \ref{fig:Parameters} can be used to set a desired bandgap and selectively place a TR based on the boundary condition of choice.}

\vspace{0.25cm}

\begin{figure*}[ht]
     \centering
\includegraphics[width=\linewidth]{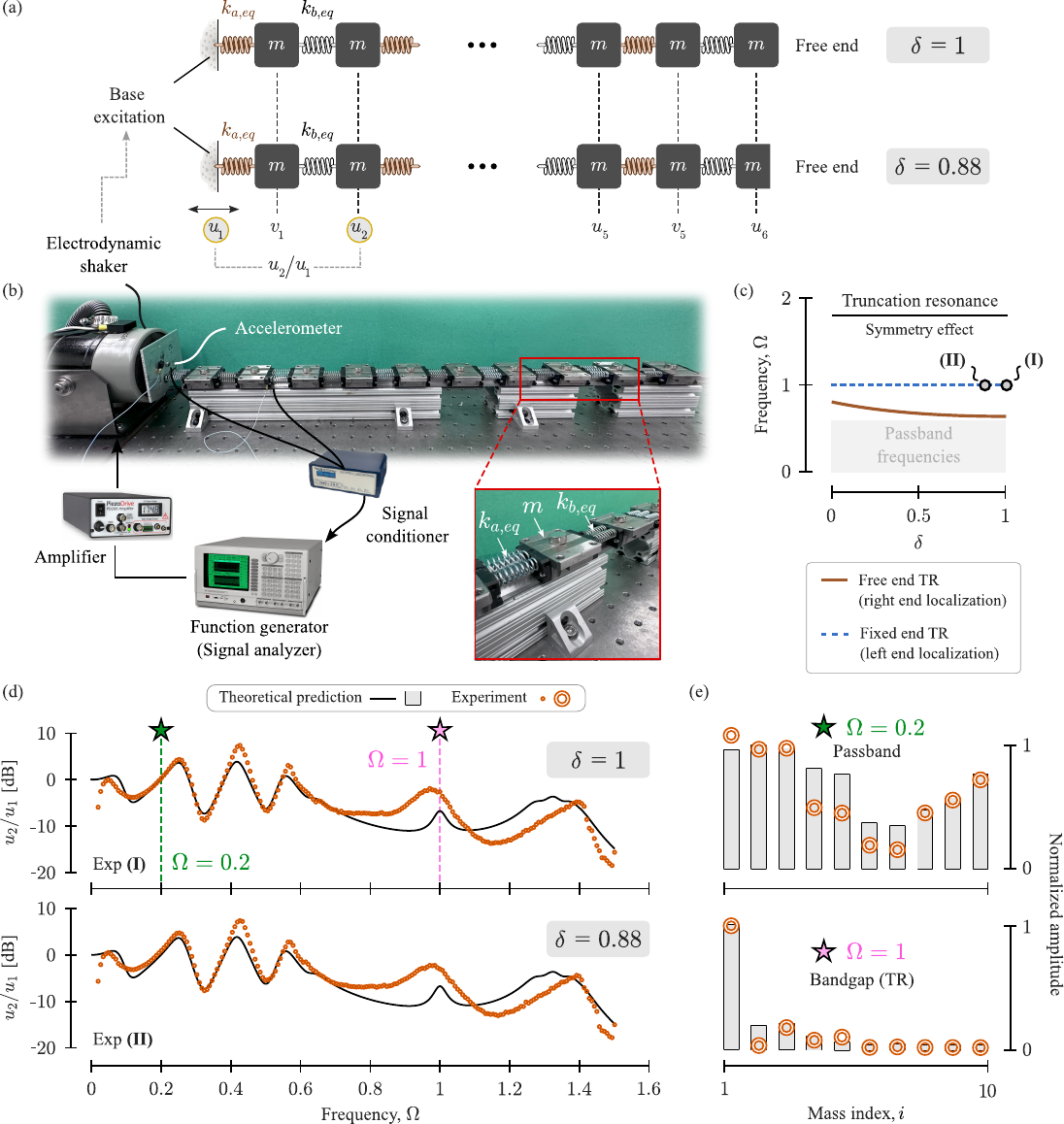}
    \begin{singlespace}
     \caption{\textbf{Experimental observation of TR.} (a) Schematic diagram of a finite diatomic lattice with fixed-free BCs. (b) Photograph and outline of the experimental setup. A finite spring-mass system is mounted to an electrodynamic shaker and is free to oscillate at the opposing end. The system consists of five diatomic unit cells that slide on a horizontal rail, each consisting of two identical masses and alternating springs, such that $\varrho = 0$ and $\vartheta = -0.59$. (c) Unit cell symmetry effect on the TRs for the chosen parameters. (d) The frequency response function $u_2/u_1$ is measured for two systems corresponding to $\delta = 1$ (top; Exp \textbf{I}) and $\delta = 0.88$ (bottom; Exp \textbf{II}) showing good agreement between the theoretical model and experiment with a TR at $\Omega = 1$. (e) Spatial wave profiles at $\Omega = 0.2$ (top) and $\Omega = 1$ (bottom) for the case when $\delta = 1$. The bottom profile confirms the left (fixed) end energy localization associated with the TR falling within the Bragg bandgap, as predicted from the analytical model.}
     \label{fig:Experiment}
    \end{singlespace}
\end{figure*}

\textbf{Experimental realization of TR.}~The setup shown in Fig.~\ref{fig:Experiment} depicts a spring-mass chain that is hooked to an electrodynamic shaker and is free to oscillate at the opposing end. The setup provides experimental evidence of the emergence of a TR within the system's frequency spectrum, which falls inside the bandgap consistent with the aforementioned analytical framework. The system shown is comprised of five diatomic unit cells. Each unit cell consists of two identical carts that slide on a horizontal rail, as shown in Fig.~\ref{fig:Experiment}(b). Each cart is $0.17$ kg and has an add-on $0.01$ kg magnet, as shown in the figure, which results in $m_{a,b} = m = 0.18$ kg, corresponding to $\varrho = 0$ from Eq.~(\ref{eq:varrho}). The add-on magnets allow us to test the system's performance for different values of $\delta$ as shown in Fig.~\ref{fig:Experiment}(a). The masses are connected using two sets of parallel springs, the equivalent stiffness of which amounts to two alternating spring constants, $k_{a,eq} = 25,086$ N/m and $k_{b,eq} = 6,455$ N/m, such that $\vartheta = -0.59$ according to Eq.~(\ref{eq:vartheta}). Using Eq.~(\ref{eq:w0}), this set of inertial and stiffness parameters yields a harmonic mean of $\omega_0 = 418.6$ rad/s. 

The system is subjected to base excitation from the shaker over a swept frequency range spanning $0 \leq \Omega \leq 1.5$.~A pair of accelerometers are placed on the end connected to the shaker and the first mass of the second unit cell to measure the frequency response function $u_2/u_1$ of the finite diatomic chain. To account for losses in the experimental setup, the theoretical equations of motion are augmented with a Rayleigh damping matrix with a damping ratio of $\zeta\approx 0.015$.~Figure~\ref{fig:Experiment}(c) shows the TR variation with the symmetry parameter $\delta$ for the chosen experimental parameters. A dispersion analysis of the same system reveals a bandgap between $\Omega=0.6$ and $1.3$. Figure~\ref{fig:Experiment}(d) compares the experimentally-obtained frequency response function with the theoretical prediction for two different configurations corresponding to $\delta = 1$ (top; Exp \textbf{I}) and $\delta = 0.88$ (bottom; Exp \textbf{II}). The second configuration was achieved by removing the magnet from the end mass yielding a reduced end mass of $0.17$ kg, as depicted in the lower sketch of Fig.~\ref{fig:Experiment}(a). The results show very good agreement between theory and experimental measurements of the transfer function $u_2/u_1$, and confirm the independence of the TR at $\Omega = 1$ on $\delta$, as indicated by the blue dashed line in Fig.~\ref{fig:Experiment}(c). The second TR corresponding to the free-end localization, and preceding the TR at $\Omega = 1$, is shown to be heavily attenuated in both the theoretical and experimental values of $u_2/u_1$. Finally, the displacement magnitudes of each mass in the finite chain are experimentally measured at a passband frequency ($\Omega = 0.2$) as well as at the TR ($\Omega = 1$) to showcase the difference in the spatial wave profile at both instances. The TR at $\Omega =1$ corresponds to a left-end localization of vibrational energy. Figure~\ref{fig:Experiment}(e) shows the experimental vs theoretical normalized displacement amplitudes vs the mass index $i$ for both cases, clearly showing the attenuating mode at the TR frequency. The shown profile also confirms the localization at the left end of the chain (where the system is anchored to the shaker), as theoretically predicted, and validates the usage of the analytical framework as a blueprint for TRprediction in elastic diatomic lattices with unit cell asymmetry.

\vspace{0.25cm}

\textbf{Concluding remarks.} We provided a robust analytical theory for the prediction of TRs emerging within the bandgaps of elastic lattices which consist of repeating monatomic and diatomic unit cells. The transfer matrix method was employed as a tool to concurrently evaluate the response of an infinite and finite elastic lattice through the dispersion and frequency response functions, respectively.~It was shown that unit cell symmetry (quantified by the parameter $\delta$), boundary conditions, and mass and stiffness contrasts ($\varrho$ and $\vartheta$), all play a critical role in the placement and location of TRs. Specifically, we demonstrated that a monatomic lattice is limited to one TRin the unbounded bandgap region, which ceases to exist when the lattice exhibits fixed boundaries or if the finite system is comprised of an integer number of fully symmetric unit cells (i.e., $\delta =0$). On the other hand, two TRs are associated with a diatomic lattice with free boundaries, one inside the Bragg bandgap while the other appears in the unbounded stop band. The framework was supported via experiments, and a set of maps was provided that encapsulate the effect of different design parameters and enable the realization and accurate placement of TRs in such lattices.

\vspace{0.25cm}

\textbf{Acknowledgements.} The authors acknowledge support of this work by the US National Science Foundation through research awards nos.~1847254 (CAREER) and 1904254, and the US Air Force Office of Scientific Research through award no. FA9550-23-1-0564.

\bibliography{references}

\end{document}